\documentclass[showpacs,showkeys]{revtex4}%
\usepackage[colorlinks=true]{hyperref}

\AtBeginDocument{% optionally set colors to your liking
	\hypersetup{
		urlcolor=cyan,
		citecolor=magenta,
		linkcolor=orange,
		anchorcolor=green,
	}%
}
\usepackage{amssymb}
\usepackage{amsmath}
\usepackage{amsfonts}
\usepackage{graphicx}
\usepackage{ulem} %\sout{strike out}

\usepackage{cancel}%\cancelto{0}{x}
\usepackage[usenames]{xcolor}
\usepackage{epstopdf}
\usepackage{extarrows}
\providecommand{\sqbr}[1]{\left[ #1 \right]} %

\def\kB{k_{\text{\tiny{B}}}}

\begin{document}

\title[ ]{Opening Pandora's Box: Maximizing the $q$-entropy with Escort Averages}
\author{$^{1}$Aruna Bidollina}
\author{$^{2}$Thomas Oikonomou}
\email{thomas.oikonomou@nu.edu.kz}
\author{$^{3}$G. Baris Bagci}
\affiliation{$^{1}$Department of Mathematics, School of Science and Technology, Nazarbayev University, Astana 010000, Kazakhstan}
\affiliation{$^{2}$Department of Physics, School of Science and 	Technology, Nazarbayev University, Astana 	010000, Kazakhstan}
\affiliation{$^{3}$Department of Physics, Mersin University, 33110 Mersin, Turkey}

\begin{abstract}
It is currently a widely used practice to write the constraints in terms of escort averages when the generalized entropies are employed in the maximization scheme. We show that the maximization of the nonadditive $q$-entropy with escort averages leads either to an overall lack of connection with thermodynamics or violation of the second and third laws of thermodynamics if one adopts the Clausius definition of the physical temperature. If an alternative definition of physical temperature is chosen by respecting  the divisibility of the total system into independent subsystems, thermodynamic relations are restored albeit at the cost of transforming the nonadditive $q$-entropy into the R\'enyi entropy. These results are illustrated by studying the quantum mechanical free particle.
\end{abstract}

\eid{ }
\date{\today }
\startpage{1}
\endpage{1}
\maketitle

%\newpage
%==================================================
\section{Introduction}
%==================================================
%
The nonadditive $q$-entropy (historically called as the nonextensive entropy) \cite{Tsallis1,Tsallis2}, despite its numerous applications in many diverse fields \cite{Misra,Behery,Nobre,Rouse,Cirto,Bagci1,Singh,Ponmurugan,Saberian,Guha,Parvan,Bagci2}, still presents some open problems such as its connection to thermodynamics and the definition of a physical temperature \cite{Abe1,Abe2}, stability of the averaging schemes \cite{Abe3,Lutsko} and its axiomatic foundations \cite{Presse1,Presse2}. One such issue still under debate is how the entropy maximization is to be carried out \cite{Jaynes,Mendes,Oikonomou1}. In particular, this issue revolves around the averaging scheme one should adopt concerning the internal energy constraint if one aims to have a consistent equilibrium distribution and associated thermodynamical structure.

Historically, the internal energy in nonextensive theory was defined as usual i.e. $U = \sum_{i}p_i\varepsilon_i$ in terms of the ordinary linear averaging scheme \cite{Mendes}. Having understood it to be problematic for various reasons, a second choice has been the expression $U= \sum_{i}p_i^q\varepsilon_i$ for the internal energy in the functional associated with the entropy maximization (MaxEnt) procedure. Due to its severe drawbacks such as violation of the energy conservation for example, the third and so far final choice has been the so-called escort averaged internal energy expression which reads $U_q= \frac{\sum_{i}p_i^q\varepsilon_i}{\sum_k p_k^q}$ \cite{Mendes}.

However, the issue of the escort averaged internal energy presents two distinct problems. First, there emerges two different expressions of the equilibrium distribution and its concomitant partition function even though one uses the same constraint for the internal energy i.e. $U_q= \frac{\sum_{i}p_i^q\varepsilon_i}{\sum_k p_k^q}$ \cite{Mendes}. Therefore, any discussion of the $q$-entropy maximization with the escort averages should consider both of these distributions to assess the feasibility of this type of maximization.
Second problem is the vagueness of the physical (inverse) temperature when one employs the escort constraints \cite{Abe1,Abe2}. If one assumes the divisibility of the total system into its subsystems despite the nonadditivity of the $q$-entropy, then one is forced to use $\frac{\beta}{\sum_{i} p_{i}^{q}}$ ($\beta$ being the Lagrange multiplier associated with the internal energy constraint) as the physical (inverse) temperature for thermodynamic inconsistency. On the other hand, if one considers the Clausius entropy as the point of departure, one should instead simply use $\beta$ as the physical (inverse) temperature \cite{Abe1,Abe2}. Therefore, one should consider both approaches to the physical temperature to ensure a correct assessment of the issue.

Our aim in this work is to attempt a detailed study of the nonadditive $q$-entropy maximization with escort averaged internal energy expression. We will consider alternative forms of the MaxEnt probability distribution and also two distinct physical temperature expressions. The prevalent inconsistencies thereby found are also illustrated through the free particle model. Concluding remarks are presented in section III.      

%============================================================================
\section{The nonadditive $q$-entropy, Escort averages and thermodynamics}
%============================================================================

The functional to be maximized reads 
\begin{eqnarray}
\Phi =\frac{S_q}{\kB} - \alpha\sqbr{\sum_i p_i -1} -\beta \sqbr{\frac{\sum_{i}p_i^q\varepsilon_i}{\sum_k p_k^q} - U_q}\,,
\end{eqnarray}
where  $S_q=\kB\sum_i p_i \ln_q(1/p_i)$ is the nonadditive $q$-entropy written in terms of the $q$-deformed logarithm $\ln_{q} (x) = \frac{x^{1-q}-1}{1-q}$ \cite{Oikonomou2}, $U_q$ is the internal energy and $\kB$ denotes the Boltzmann constant. Taking the partial derivative of the above functional with respect to $p_i$ and then equating it to zero, we obtain 
\begin{eqnarray}\label{eq2}
\frac{qp_i^{q-1}}{1-q}\sqbr{1 - (1-q)\frac{\beta}{\sum_k p_k^q}(\varepsilon_i -U_q)}-\frac{1}{1-q}-\alpha&=&0
\end{eqnarray}
which can be cast into the form below
\begin{eqnarray}
p_i=\sqbr{\frac{1+(1-q)\alpha}{q}}^{\frac{1}{q-1}}\sqbr{ 1 - (1-q)\frac{\beta}{\sum_k p_k^q}(\varepsilon_i -U_q)}^{\frac{1}{1-q}}\,.
\end{eqnarray}
Probability normalization yields then
%
%\begin{eqnarray}\label{eq6a}
%\sqbr{\frac{1+(1-q)\alpha}{q}}^{\frac{1}{1-q}}=\sum_i \sqbr{ 1 - (1-q)\frac{\beta}{\sum_k p_k^q}(\varepsilon_i -U_q)}^{\frac{1}{1-q}}
%\end{eqnarray}
%
%so that the equilibrium distribution associated with the escort averaged internal energy reads
%
\begin{eqnarray}\label{MaxEntDistr}
p_i=\frac{1}{Z_q} \sqbr{ 1 - (1-q)\frac{\beta}{\sum_k p_k^q}(\varepsilon_i -U_q)}^{\frac{1}{1-q}}
\end{eqnarray}
with the $q$-partition function $Z_q$ is given as
\begin{eqnarray}\label{eq6}
Z_q =\sqbr{\frac{1+(1-q)\alpha}{q}}^{\frac{1}{1-q}}=\sum_i \sqbr{ 1 - (1-q)\frac{\beta}{\sum_k p_k^q}(\varepsilon_i -U_q)}^{\frac{1}{1-q}}\,.
\end{eqnarray}

If one multiplies Eq. (\ref{eq2}) with $p_i$ and sums over the all $i$'s, one obtains
%
%\begin{eqnarray}
%\sum_i p_i^{q} = \frac{1}{q}\sqbr{1+(1-q)\alpha}\,.
%\end{eqnarray}
%
%Comparing the above equation with Eq. (\ref{eq6}), we obtain the following important relation 
%
\begin{eqnarray}\label{keyEq}
\sum_i p_i^q = (Z_{q})^{1-q}\,,
\end{eqnarray}
where Eq. (\ref{eq6}) is taken into account.
Thus, the nonadditive $q$-entropy of the equilibrium distribution is calculated as
\begin{eqnarray}
S_q=\kB\sum_i p_i \ln_q(1/p_i)=\kB\frac{\sum_i p_i^q -1}{1-q}=\kB\frac{(Z_q)^{1-q}-1}{1-q}
\end{eqnarray}
which, in terms of the deformed $q$-logarithm, reads
\begin{eqnarray}\label{main1}
S_q=\kB\ln_q(Z_{q}).
\end{eqnarray}
This is an important result worth pondering. First of all, the expression above implies that the essential link between the statistical mechanics and thermodynamics is severely missing, since one cannot construct a relation between the entropy, partition function and the average energy but only between the former two. As a result, one cannot trust a consistent connection to exist between the statistical mechanics and thermodynamics based on the nonadditive $q$-entropy if one employs the escort averaged internal energy constraints. Note that it has recently been shown that the R\'enyi entropy is equal to the (natural) logarithm of the partition function when one uses the ordinary internal energy definition so that it has been concluded that there is no R\'enyi thermodynamics (see Ref. \cite{Plastino} and in particular Eq. (4.15) therein).

Second, the inspection of Eq. (\ref{eq6}) in the $q\to1$ limit shows that one then has $Z_{1} = e^{\beta U_1} \sum_{i} e^{- \beta \varepsilon_i} $  for the partition function in this particular limit. We recall that one very general feature of the nonextensive theory as a generalization scheme is that one should obtain the expressions in ordinary statistical mechanics whenever this limit is invoked. A comparison of $Z_1$ with the  ordinary canonical partition function $Z= \sum_{i} e^{- \beta \varepsilon_i}$ explicitly shows that they are different from one another, the former including a multiplicative $e^{\beta U_1}$ term i.e. $Z_1 = e^{\beta U_1} Z$. Note, however, that the cancellation of this extra term in both nominator and denominator of Eq. (\ref{MaxEntDistr}) in the limit $q=1$ enables one to obtain the ordinary canonical distribution \cite{note1}.  In other words, employing escort averaged internal energy constraint yields a generalized partition function which does not warrant the textbook canonical partition function in the appropriate limit.

In order to illustrate the viewpoint above, we now consider a quantum mechanical free particle \cite{Caldeira}. The density operator for the $q$-entropy with escort distributions is given by
\begin{eqnarray}\label{density}
\hat{\rho}_q=\frac{\hat{A}_q}{Z_q(\beta)}\,,\qquad \hat{A}_q:=\left[1-(1-q)\beta_q(\hat{H}-U_q)\right]^{\frac{1}{1-q}}\,,
\end{eqnarray}
where we confine ourselves only to $q>1$ interval \cite{OikGBB2009,OikGBB2010}. The partition function $Z_q$ and the energy factor $\beta_q$ are given by
\begin{eqnarray}
Z_q(\beta)=\mathrm{Tr}\left\{\hat{A_q} \right\} \,, \qquad \qquad 
\beta_q:= \frac{\beta}{\mathrm{Tr}\left\{(\hat{\rho}_q)^q\right\}}=\frac{\beta}{[Z_q(\beta)]^{1-q}} \,.
\end{eqnarray}
Using the integral representation of $Z_q$ \cite{Mendes2}, we can write it as
\begin{eqnarray}
Z_q(\beta)  = \frac{1}{\Gamma\left(\frac{1}{q-1}\right)}\int_0^\infty \mathrm{d}t\,e^{-t[1+\beta_q(q-1)(U_{1}-U_{q})]}Z_1[t\beta_q(q-1)]t^{\frac{1}{q-1}-1}\,,
\end{eqnarray}
where
\begin{eqnarray}\label{Eq03}
Z_1(\beta)=\mathrm{Tr}\left\{ e^{-\beta(\hat{H}-U_1)}\right\}.
\end{eqnarray}

In particular, for the free particle of mass $m$ confined in one-dimensional length $L$, we have $Z_1 (\beta)=e^{\beta U_1} L\left(\frac{m}{2\pi \beta \hbar^2}\right)^{\frac{1}{2}}$ so that its substitution into the equation above yields
\begin{eqnarray}\label{KeyEq}
Z_q(\beta)&=& (\beta_q)^{-\frac{1}{2}} \widetilde{L}_q\left[1-(q-1)\beta_q U_q \right]^{\frac{1}{1-q}+\frac{1}{2}}\,,\qquad  
 \widetilde{L}_q:= L\left(\frac{m}{2\pi \hbar^2}\right)^{\frac{1}{2}} \frac{\Gamma\left(\frac{1}{q-1}-\frac{1}{2}\right)}{\Gamma\left(\frac{1}{q-1}\right) \sqrt{q-1}}
\end{eqnarray}
with $q\in(1,3)$, $\beta_q U_q<1/(q-1)$ and $\widetilde{L}_1=L\left(\frac{m}{2\pi \hbar^2}\right)^{\frac{1}{2}}$. The substitution of the relation $\beta_q=\frac{\beta}{[Z_q(\beta)]^{1-q}} $ into the equation above gives
\begin{eqnarray}\label{Eq13}
[Z_q(\beta)]^{\frac{1+q}{2}}&=& \beta^{-\frac{1}{2}}\widetilde{L}_q \left[1-(q-1)\frac{\beta}{[Z_q(\beta)]^{1-q}} U_q\right]^{\frac{1}{1-q}+\frac{1}{2}}\,.
\end{eqnarray}
%
%where
%
%\begin{eqnarray}\label{Eq13b}
% \widetilde{L}_q:= L\left(\frac{m}{2\pi \hbar^2}\right)^{\frac{1}{2}} \frac{\Gamma\left(\frac{1}{q-1}-\frac{1}{2}\right)}{\Gamma\left(\frac{1}{q-1}\right) \sqrt{q-1}}
%\end{eqnarray}
%
Solving Eq. (\ref{Eq13}) with respect to $U_q$ we have
\begin{eqnarray}\label{Eq14}
U_q(\beta)=\frac{[Z_q(\beta)]^{1-q}}{\beta} \ln_q\left(\frac{[Z_q(\beta)]^{\frac{1+q}{3-q}}\beta^{\frac{1}{3-q}}}{[\widetilde{L}_q]^{\frac{2}{3-q}}}\right).
\end{eqnarray}
Calculating the derivative with respect to $\beta$ and taking into account the relation $\frac{\partial}{\partial\beta}\ln_q(Z_q)=\beta\frac{\partial}{\partial\beta} U_q$ we have
\begin{eqnarray}
\frac{1}{\beta}\frac{\partial \ln_q(Z_q(\beta))}{\partial \beta}=\frac{\partial}{\partial\beta}\left\{\frac{[Z_q(\beta)]^{1-q}}{\beta} \ln_q\left(\frac{[Z_q(\beta)]^{\frac{1+q}{3-q}}\beta^{\frac{1}{3-q}}}{[\widetilde{L}_q]^{\frac{2}{3-q}}}\right)\right\}.
\end{eqnarray}
After some simple algebra, this differential equation reduces to 
\begin{eqnarray}\label{diffEq}
	\left[Z_q(\beta)-2(1-q)\beta\frac{\partial Z_q(\beta)}{\partial \beta} \right] \left[q-3+2 \left(\frac{\beta}{\widetilde{L}_q^2}\right)^{\frac{1-q}{3-q}}[Z_q(\beta)]^{\frac{(1-q)(1+q)}{3-q}}  \right]=0.
\end{eqnarray}
As can be seen here there are two distinct solutions of the differential equation in Eq. (\ref{diffEq}), namely
\begin{subequations}\label{Eq16}
\begin{eqnarray}
	Z^{(1)}_q(\beta) &=& c[2(1-q)]^{\frac{1}{2(1-q)}}\;\beta^{\frac{1}{2(1-q)}}\,, \\
	Z^{(2)}_q(\beta)&=&\left(\widetilde{L}_q\right)^{\frac{2}{1+q}}\left(\frac{3-q}{2}\right)^{\frac{3-q}{1-q^2}}\beta^{-\frac{1}{1+q}}\,,
\end{eqnarray}
\end{subequations}
where $c$ is the integration constant.

This is indeed a very strange situation, since the partition function uniquely describes a physical system. In other words, it does not make sense to have two partition functions for the same physical system. The first solution does not converge in the $q\to1$ limit, and therefore can be discarded. It is very important though to understand that our assessment above has been solely based on a form of mathematical argument which relies on the assurance of the ordinary thermostatistics i.e. the existence of a canonical distribution in the $q\to1$ limit. Left to the nonextensive theory by itself, there would be no way to choose between these two partition functions on their own merits. Since we know that there exists a canonical partition function for a free particle from the ordinary thermostatistics, we deduce that the limit $q\to1$ should exist and only the second partition function above i.e. $Z^{(2)}_q(\beta)$ conforms to this requirement. Thus, for the sake of simplicity, we will denote $Z_q^{(2)}$ as $Z_q$. However, there now emerges another problem, since the partition function $Z_q$  does not yield the correct canonical expression in the $q\to1$ limit although it surely exists, namely
\begin{eqnarray}\label{ErrPF}
	Z_1(\beta)=\sqrt{e} \;Z(\beta).
\end{eqnarray}
In fact, this is to be expected, since the ordinary average internal energy for the free particle is $U_1=U= \frac{1}{2 \beta}$ so that the aforementioned relation $Z_1 = e^{\beta U_1} Z (\beta)$ yields Eq. (\ref{ErrPF}). As previously explained, this is an artefact of severing the link between thermodynamics and statistical mechanics as a result of employing escort averaged internal energy expression i.e. Eq. (\ref{main1}).

Nevertheless, one can avoid this problem by rewriting the density operator in Eq.  (\ref{density}) as
\begin{eqnarray}\label{NDistr}
	\hat{\rho}_q=\frac{[1-(1-q)\overline{\beta}_q \hat{H}]^{\frac{1}{1-q}}}{\overline{Z}_q(\beta)}\,,
\end{eqnarray}
where 
\begin{eqnarray}\label{eq21}
	\overline{Z}_q(\beta)=\frac{Z_q(\beta)}{\exp_q(\beta_q U_q)}\qquad\text{and}\qquad \overline{\beta}_q=\frac{\beta_q}{1+(1-q)\beta_q U_q}.
\end{eqnarray}
Now, it can be easily checked that $\overline{Z}_q(\beta)$ yields the correct canonical expression in the $q\to1$ limit. Thus, one might be inclined to adopt $\overline{Z}_q$ to be the ultimate partition function associated with the escort averaging procedure.

\begin{figure}[h]
	\centering
	\includegraphics[keepaspectratio,width=10.0cm]{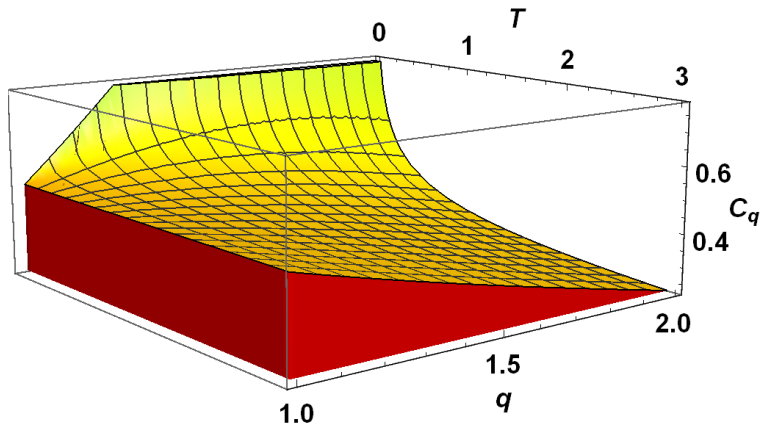}
	\caption{The heat capacity $C_q$ as a function of the temperature $T$ for the interval $q\in(1,2)$.}
	\label{fig1}
\end{figure}

Despite this succesful rewritting, another difficulty emerges now, since one cannot be sure of how thermodynamic observables will behave under this rewriting. We certainly know that these new expressions will correctly yield the respective canonical expressions, but we have no clue whatsoever on whether they will exhibit consistent behavior for all $q$ values except $q\to 1$ i.e. the ordinary canonical limit. As a case in point, we reconsider the free particle example and obtain 
\begin{subequations}\label{MEn}
\begin{eqnarray}
\label{MEn_a}
	\beta_q &=& \left(\widetilde{L}_q\right)^{\frac{2(q-1)}{1+q}}\left(\frac{3-q}{2}\right)^{\frac{q-3}{q+1}}\beta^{\frac{2}{1+q}}\,,\\
	\overline{\beta}_q&=&\frac{2}{3-q}\beta_q\,,\qquad \overline{Z}_q(\beta)=\frac{Z_q(\beta)}{\exp_q(1/2)}\,,\\
\label{MEn_c}	U_q&=&\frac{1}{2}(\widetilde{L}_q)^{\frac{2(1-q)}{1+q}}\left(\frac{3-q}{2}\right)^{\frac{3-q}{q+1}}\beta^{-\frac{2}{1+q}}=\frac{1}{2\beta_q}\,.
\end{eqnarray}
\end{subequations}
It can be checked that all these quantities correctly recover the respective canonical ones. One can also calculate the heat capacity using the above expressions so that one has
\begin{eqnarray}
	C_q=\frac{\partial U_q}{\partial T}=\frac{\kB }{1+q}(\widetilde{L}_q)^{\frac{2(1-q)}{1+q}}\left(\frac{3-q}{2}\right)^{\frac{3-q}{q+1}} \;(\kB T)^{\frac{1-q}{1+q}}\,,
\end{eqnarray}
where $\kB T=\beta^{-1}$. In Fig. 1, for the interval $q\in(1,2)$, we plot this heat capacity $C_q$ as a function of both temperature and the non-additivity parameter $q$ where we set $L=m=\hbar^2=\kB=1$. It can be seen that the heat capacity $C_q$ decays for increasing temperature therefore violating both 2nd and 3rd laws of thermodynamics. Therefore, the rewriting of the distribution given in Eqs. (\ref{NDistr}) and (\ref{eq21}) is unphysical. Note that the same unphysical behavior of the heat capacity occurs for the same model even though one adopts the ordinary internal energy expression (see Fig. 5 and related explanations in Ref. \cite{Caldeira}). To sum up, the maximization of the nonadditive $q$-entropy with the escort averages, depending on how we choose to write the resulting equilibrium distribution, either severs the link between the statistical mechanic and thermodynamics (see Eq. (\ref{main1}) and explanations below it) and does not yield the ordinary canonical partition function in the appropriate limit or results in the violation of the second and third laws of thermodynamics (see Fig. \ref{fig1}).   

In fact, one may even say that this happens because $T$ is not the physical temperature in nonextensive systems. Note that there are two definitions of (inverse) temperature in the literature when the escort averages are used, one stemming from the Clausius relation \cite{Abe2} and the other from the assumption of divisibility of the total system into independent subsystems \cite{Abe1}. The former is the Lagrange multiplier $\beta$ and corresponds to the (inverse) temperature definition we adopted so far. The latter is defined as $\kB T_q^{\text{phys}}:=(\beta_q)^{-1}$ so that the internal energy and the heat capacity become
\begin{eqnarray}\label{HCap}
U_q=\frac{\kB T_q^{\text{phys}}}{2} \qquad\Rightarrow\qquad 
C_q=\frac{\partial U_q}{\partial T_q}=\frac{\kB}{2}\,.
\end{eqnarray}
We now have both the consistent limits and sensible behaviors for the thermodynamic observables.

Despite this improvement though, we face two novel and serious drawbacks. First drawback is that the thermodynamic observables such as internal energy and heat capacity have now exactly the same form as in the ordinary canonical case. As a result, it is not clear at all why one should use the nonextensive theory instead of the ordinary canonical scheme. In other words, as far as the thermodynamic observables are concerned, nonextensive theory seems redundant although the underlying, unobservable entropy $S_q$ may be different from the ordinary Boltzmann-Gibbs-Shannon entropy.

Second and more serious drawback can be noted by realizing that the change of the inverse temperature $\beta$ to $\beta_q$ used above is not merely a substitution but a transformation implying also a transformation of the entropy expression. To see this explicitly, we begin with the following relation
\begin{eqnarray}\label{eq02}
\beta=\frac{1}{\kB}\frac{\partial S_q}{\partial U_q}=\frac{1}{\kB}\frac{\partial S_q/\partial \beta}{\partial U_q/\partial \beta}\,.
\end{eqnarray}
Taking into account Eq. (\ref{main1}) i.e. $S_q=\kB \ln_q(Z_q)$, we have
\begin{eqnarray}\label{eq03}
\frac{1}{\kB}\frac{\partial S_q}{\partial \beta} &=& 
\frac{\partial \ln_q(Z_q)}{\partial \beta}
=(Z_q)^{1-q} \frac{\partial \ln(Z_q)}{\partial \beta}\,.
\end{eqnarray}
Combining Eqs. (\ref{eq02}) and (\ref{eq03}) we have
%
%\begin{eqnarray}
%\frac{1}{\kB} \frac{\partial S_q}{\partial \beta} = \beta \frac{\partial U_q}{\partial \beta} \quad\Ra\quad
%(Z_q)^{1-q} \frac{\partial \ln(Z_q)}{\partial \beta} = \beta \frac{\partial U_q}{\partial \beta} \quad\Ra\quad
%\frac{\partial \ln(Z_q)}{\partial \beta} = \frac{\beta}{(Z_q)^{1-q}} \frac{\partial U_q}{\partial \beta} = \beta_q \frac{\partial U_q}{\partial \beta}
%\end{eqnarray}
%
%so that
%
\begin{eqnarray}
\beta_q=\frac{\partial \ln(Z_q)}{\partial U_q}.
\end{eqnarray}
The comparison of the equation above with Eq. (\ref{eq02}) shows that the entropy expression related to the temperature $\beta_q$ is not $S_q$ anymore but $S_q^\mathrm{phys}=\kB\ln(Z_q)$. Finally, using Eq. (\ref{keyEq}) for the expression $S_q^\mathrm{phys}=\kB\ln(Z_q)$, we identify this entropy as the R\'enyi entropy \cite{Renyi}
\begin{eqnarray}
S_q^\mathrm{phys}=\kB\ln(Z_q)=\frac{\kB}{1-q}\ln\left(\sum_i p_i^q\right).
\end{eqnarray}
In other words, the maximization of the entropy $S_q$ with escort constraints and the adoption of $\beta_q $ as the physical temperature are equivalent to adopting the R\'enyi entropy. Therefore, such a combination can not be solely treated in the context of the nonextensive theory any more \cite{Campisi}.

%==================================================
\section{Conclusions}
%==================================================
%

The nonadditive $q$-entropy is currently maximized by defining the internal energy as $U_q= \frac{\sum_{i}p_i^q\varepsilon_i}{\sum_k p_k^q}$ instead of the well-known expression $U = \sum_{i}p_i\varepsilon_i$. This type of averaging is called escort averaging \cite{Mendes} and it has been criticized in terms of its stability \cite{Abe3,Lutsko} before.

In this work, instead of focusing on the escort averaging \textit{per se}, we consider the maximization scheme and the resulting probability distribution. Written in one form, the distribution results in $S_q = \kB\ln_q (Z_{q})$ i.e. Eq. (\ref{main1}) without any explicit appearance of the internal energy expression. As a result, one cannot construct a bridge between statistical mechanics and thermodynamics as in the ordinary case i.e. $S/\kB=\ln Z + \beta U$. Moreover, the canonical limit of the partition function in this context is found as $e^{\beta U_1} \sum_{i} e^{- \beta \varepsilon_i} $ instead of the correct expression $\sum_{i} e^{- \beta \varepsilon_i} $. Written in another form, the distribution obtained from the escort averaged internal energy implies a violation of second and third laws of thermodynamics in thermodynamic observables as shown in Fig. \ref{fig1}. However, all these results are valid only if one adopts a physical temperature equal to the inverse of the internal energy Lagrange multiplier $\beta$. This is the definition one derives by recourse to the Clausius relation as shown in Ref. \cite{Abe2}.

On the other hand, there is another definition of physical temperature which stems from the assumption of the divisibility of the total system into independent subsystems Ref. \cite{Abe1}. When this temperature $\beta_{q}$ is considered to be the physical one, one recovers the equipartition results so that the thermodynamic consistency is restored as can be seen from Eq. (\ref{HCap}). However, when this is the case, we show that the use of this alternative physical temperature definition is tantamount to the adoption of the R\'enyi entropy, deeming the nonadditive $q$-entropy redundant. In fact, it has been previously found that the requirement of the equipartition theorem necessitates the use of the R\'enyi entropy \cite{Campisi}.

To conclude, the maximization of the nonadditive $q$-entropy with escort averages implies either thermodynamic anomalies or a transmutation of the $q$-entropy into the R\'enyi entropy, making the former redundant. Finally, note that the ordinary averaging scheme also results in thermodynamic anomalies \cite{Caldeira} or foundational inconsistencies \cite{Presse1,Presse2,Oikonomou1}. Therefore, we conclude that how constraints should be averaged is still an open problem for the nonadditive $q$-entropy.

\begin{acknowledgments}
	This research is partly supported by ORAU grant entitled ``Casimir light as a probe of vacuum fluctuation simplification" with PN 17098. T.O. acknowledges the state-targeted program ``Center of Excellence for Fundamental and Applied Physics" (BR05236454) by the Ministry of Education and Science of the Republic of Kazakhstan. G.B.B. acknowledges support from Mersin University under the project number 2018-3-AP5-3093.
\end{acknowledgments}

%============================


\begin{thebibliography}{0}
%============================


\bibitem{Tsallis1} C. Tsallis, J. Stat. Phys. \textbf{52} (1988) 479.

\bibitem{Tsallis2} C. Tsallis, \textit{Introduction to nonextensive statistical mechanics:  approaching a complex world}, New York: Springer, 2009.



\bibitem{Misra} D. Chatterjee and A. P. Misra, Phys. Rev. E \textbf{92} (2015) 063110.

\bibitem{Behery} E. E. Behery, Phys. Rev. E \textbf{94} (2016) 053205.

\bibitem{Nobre} M. S. Ribeiro and F. D. Nobre, Phys. Rev. E \textbf{94} (2016) 022120.

\bibitem{Rouse} I. Rouse and S. Willitsch, PRL \textbf{118} (2017) 143401.


\bibitem{Cirto} C.-Y. Wong, G. Wilk, L. J. L. Cirto and C. Tsallis, Phys. Rev. D \textbf{91} (2015) 114027. 

\bibitem{Bagci1} G. B. Bagci, R. Sever, and C. Tezcan, Mod. Phys. Lett. B \textbf{18} (2004) 467.

\bibitem{Singh} C.-C. Chang, R. R. P. Singh, R. T. Scalettar, Phys. Rev. B \textbf{90} (2014) 155113.


\bibitem{Ponmurugan} M. Ponmurugan, Phys. Rev. E \textbf{93} (2016) 032107.

\bibitem{Saberian} E. Saberian and A. Esfandyari-Kalejahi, Phys. Rev. E \textbf{87} (2013) 053112.


\bibitem{Guha} A. Guha and P. K. Das, Physica A \textbf{495} (2018) 18.

\bibitem{Parvan} D. V. Anghel and A. S. Parvan, J. Phys. A \textbf{51} (2018) 445002.


\bibitem{Bagci2} G. B. Bagci, Physica A \textbf{386} (2007) 79.



\bibitem{Abe1} S. Abe, S. Mart{\'i}nez, F. Pennini and A. Plastino, Phys. Lett. A \textbf{281}  (2001) 126.


\bibitem{Abe2} S. Abe, Physica A \textbf{368}  (2006) 430.

\bibitem{Abe3} S. Abe, EPL \textbf{84} (2008) 60006.

\bibitem{Lutsko} J. F. Lutsko, J.P. Boon, P. Grosfils, EPL \textbf{86} (2009) 40005.

\bibitem{Presse1} S. Press{\'e}, Phys. Rev. E \textbf{90} (2014) 052149.

\bibitem{Presse2} S. Press{\'e}, K. Ghosh, J. Lee and K. A. Dill, Phys. Rev. Lett. \textbf{111}, (2013) 180604.



\bibitem{Jaynes} E. T. Jaynes, Phys. Rev. \textbf{106} (1957) 171; \textbf{108} (1957) 620.

\bibitem{Mendes} C. Tsallis, R. S. Mendes, and A. R. Plastino, Physica A  \textbf{261} (1998) 534.

\bibitem{Oikonomou1} T. Oikonomou and G. B. Bagci, Phys. Lett. A  \textbf{381} (2017) 207.

\bibitem{Oikonomou2} G. B. Bagci and T. Oikonomou, Phys. Rev. E  \textbf{93} (2016) 022112.

\bibitem{Plastino} A. Plastino, M. C. Rocca, and F. Pennini, Phys. Rev. E  \textbf{94} (2016) 012145.


\bibitem{note1} Alternatively, one can consider Eq. (\ref{main1}) in the limit $q \to 1$ so that one has $S = \ln Z_1$ where $S$ is the ordinary Boltzmann-Gibbs-Shannon entropy i.e. $- \sum_{i} p_{i} \ln p_{i}$. The substitution of the ordinary canonical distribution into the entropy expression $S$ again yields $Z_{1} = e^{\beta U_1} \sum_{i} e^{- \beta \varepsilon_i} $ which is different than the ordinary canonical partition function expression.  

\bibitem{Caldeira} L. M. M. Dur$\tilde{a}$o and A. O. Caldeira, Phys. Rev. E  \textbf{94} (2016) 062147.

\bibitem{OikGBB2009} T. Oikonomou and G. B. Bagci, J. Math. Phys.  \textbf{50} (2009) 103301.

\bibitem{OikGBB2010} T. Oikonomou and G. B. Bagci, Phys. Lett. A  \textbf{374} (2010) 2225.

\bibitem{Mendes2} R. S. Mendes, Braz. J. Phys. \textbf{29} (1999) 66.

\bibitem{Renyi} A. R{\'e}nyi, \textit{Probability Theory}, (North-Holland) 1970.

\bibitem{Campisi} M. Campisi and G. B. Bagci, Phys. Lett. A \textbf{362} (2007) 11.

\end{thebibliography}
\end{document}